\documentclass[aps,prd,onecolumn,groupedaddress,showpacs,nofootinbib,amssymb]{revtex4}
\usepackage[dvips]{graphicx}
\usepackage{amssymb}
\usepackage{amsmath}
\usepackage{graphicx,,color}
\usepackage{amsfonts}
\usepackage{bm}
\usepackage{cancel}
\usepackage{comment}

\newcommand\be{\begin{equation}}
\newcommand\ee{\end{equation}}

\allowdisplaybreaks[4]

\begin{document}

\title{$f(R)$ Gravity Inflation with String-Corrected Axion Dark Matter}
\author{S.D. Odintsov,$^{1,2,3}$\,\thanks{odintsov@ieec.uab.es}
V.K. Oikonomou,$^{4,5,3}$\,\thanks{v.k.oikonomou1979@gmail.com}}
\affiliation{$^{1)}$ ICREA, Passeig Luis Companys, 23, 08010 Barcelona, Spain\\
$^{2)}$ Institute of Space Sciences (IEEC-CSIC) C. Can Magrans s/n,
08193 Barcelona, Spain\\
$^{3)}$ Tomsk State Pedagogical University, 634061 Tomsk, Russia\\
$^{4)}$Department of Physics, Aristotle University of Thessaloniki, Thessaloniki 54124, Greece\\
$^{5)}$ Laboratory for Theoretical Cosmology, Tomsk State University
of Control Systems
and Radioelectronics, 634050 Tomsk, Russia (TUSUR)\\
}

\tolerance=5000

\begin{abstract}
It is quite well known for some time that string inspired axionic
terms of the form $\nu (\phi)\tilde{R}R$, known also as
Chern-Simons terms, do not affect the scalar perturbations and the
background evolution for a flat Friedman-Robertson-Walker
Universe. In this paper we study and quantify the implications of
the presence of the above term in the context of vacuum $f(R)$.
Particularly, we assume that axionic dark matter is present during
inflation, and we examine in a quantitative way the effects of
axionic Chern-Simons terms on the tensor perturbations. The axion
field is quantified in terms of a canonical scalar field, with
broken Peccei-Quinn symmetry. The model perfectly describing
axions as potential dark matter candidates is based on the
so-called misalignment mechanism, in which case the axion is
frozen near its non-zero vacuum expectation value during early
times with $H\gg m_a$. In effect, the inflationary era is mainly
controlled by the $f(R)$ gravity and the Chern-Simons term. As we
demonstrate, the Chern-Simons term may achieve to make a
non-viable $f(R)$ gravity theory to be phenomenologically viable,
due to the fact that the tensor-to-scalar ratio is significantly
reduced, and the same applies to the spectral index of the tensor
perturbations $n_T$. Also by studying the Starobinsky model in the
presence of the Chern-Simons term, we demonstrate that it is
possible to further reduce the amount of primordial gravitational
radiation. The issues of having parity violating gravitational
waves, also the graceful exit from inflation due to axion
oscillations and finally the unification of dark energy-inflation
and axion dark matter in the same $f(R)$ gravity-axion dark matter
model, are also briefly discussed.
\end{abstract}


\maketitle

\section{Introduction}

Cosmology is currently at a stage in which many observational data
exist that need to be appropriately explained theoretically. Three
are the most intriguing mysteries in cosmology, the dark matter
nature, the dark energy issue for the late-time Universe, and the
primordial era. These mysteries are currently constrained by
observations, but for the moment we are at the stage of
speculating and fitting models that may appropriately describe
these in a consistent way. With regard to the primordial era, it
is currently accepted that an inflationary era preceded the
radiation and matter domination eras, and many theoretical
proposals have appeared after the pioneer papers
\cite{Guth:1980zm,Starobinsky:1982ee,Linde:1983gd} that  can
describe this accelerating era. The inflationary era and the dark
energy epoch may be successfully described by modified gravity
\cite{reviews1,reviews2,reviews3,reviews4,reviews5,reviews6}, and
actually a unified description can be provided in terms of a
modified gravity, see for example \cite{Nojiri:2003ft} for an
early $f(R)$ gravity model unifying dark energy and inflation.

Dark matter was introduced in order to describe theoretically the
galactic rotation curves, and its nature is mysterious. Although
modified gravity may explain to some extent certain aspects of
dark matter, after the $GW170817$ event
\cite{TheLIGOScientific:2017qsa}, these theories were seriously
questioned \cite{Boran:2017rdn}. Thus the only consistent for the
moment theoretical proposal is that dark matter is some weakly
interacting massive particle. In fact, one fascinating event that
strongly supports the particle nature of dark matter is collision
of galaxies observed in the Bullet Cluster, which indicates that
there is an invisible gravitating component in the collision. In
the literature there exist many candidate particles that may
describe dark matter, see for example \cite{Oikonomou:2006mh}, but
still no direct detection of dark matter occurred. One promising
candidate is the axion \cite{Marsh:2015xka,Marsh:2017yvc}, which
occurs in various theoretical physics contexts, even in string
theory, see also \cite{maxim} for some related works on axion dark
matter. The axion is a low-mass particle which is the goldstone
boson of the spontaneously broken Peccei-Quinn symmetry, in the
context of ordinary field theory and also in string theory. In
string theory the axions are massless when the extra dimensional
theory is compactified in four dimensions, and non-perturbative
effects can provide mass to axions. The most interesting model for
axion dark matter is the so-called misalignment axion model, in
which initially the axion has a small and constant mass during and
before the inflationary era, and starts to oscillate during the
radiation domination era. In this paper we shall be interested in
investigating the effects of the existence of axions in an $f(R)$
gravity inflationary theory. We shall assume that the axion is
described by a canonical scalar field at early times and also that
the axion field is a misalignment axion and it thus has a non-zero
vacuum expectation value. Also we shall assume that terms coming
from string corrections are present in the theory, of the form
$\nu (\phi)\tilde{R}R$. These are well motivated from string
theory, due to the fact that string theory is the only consistent
UV description of all known theories describing the four
fundamental interactions. It is thus possible that some terms may
survive during the inflationary era, and these can have some
effect on the dynamics of inflation. Terms of the form $\nu
(\phi)\tilde{R}R$ are dubbed Chern-Simons (CS) terms, however the
term $\nu (\phi)\tilde{R}R$ is simply the Chern-Pontryagin density
which actually is related to a three dimensional Chern-Simons term
via the exterior derivative $\nu
(\phi)\tilde{R}R=d(Chern-Simons)$. An important stream of papers
on Chern-Simons gravity can be found in Refs.
\cite{Nishizawa:2018srh,Wagle:2018tyk,Yagi:2012vf,Yagi:2012ya,Molina:2010fb,Izaurieta:2009hz,Sopuerta:2009iy,Konno:2009kg,Smith:2007jm,Matschull:1999he,Haghani:2017yjk,Satoh:2007gn,Satoh:2008ck,Yoshida:2017cjl}
and references therein. The effect of the CS term on theories of
the form $f(R,\phi)$ were thoroughly studied by Hwang and Noh some
time ago \cite{Hwang:2005hb} and as they proved, the CS term
affects the tensor perturbations and leaves intact the scalar
perturbations and the background evolution. In this paper we are
interested to quantify the effect of the presence of an axion dark
matter, with a CS term present, on vacuum $f(R)$ gravity theories.
Our aim is two-fold: Firstly we choose some $f(R)$ gravity which
is known to provide a non-viable phenomenology, and try to
investigate whether the axion dark matter with CS term can make
the theory viable. Secondly we shall choose the most successful
model of $f(R)$ gravity up-to-date, namely the Starobinsky model
\cite{Starobinsky:1982ee}, and we shall investigate quantitatively
the effect of CS corrected axion dark matter on the vacuum $f(R)$
gravity. As we shall demonstrate, the effect of the CS term is to
reduce the value of tensor-to-scalar ratio and of the spectral
index of the primordial tensor perturbations, and we provide a
thorough quantitative analysis of how these reductions can be
produced. The resulting picture of the $f(R)$-dark matter model we
introduce is quite interesting theoretically, since the presence
of early-time dark matter affects the dynamics of inflation, and
also has a non-trivial effect on the propagation models of
gravitational waves, discriminating the two different
polarizations.

This paper is organized as follows: In section II we shall discuss
in some detail the CS modified axion dark matter $f(R)$ gravity
model, focusing also on the essential properties of the axion
misalignment model. In section III we shall investigate the
implications of the CS axion dark matter on a power-law $f(R)$
gravity which is known to provide a non-viable inflationary
phenomenology in vacuum. Particularly we shall investigate when
the resulting model can be compatible with the Planck
\cite{Ade:2015lrj} and BICEP2/Keck-Array \cite{Array:2015xqh}
observational data. In addition, we examine the effects of the CS
modified axion dark matter on the Starobinsky model, and finally
the conclusions follow in the end of the paper.

Before we start off, until it is stated differently, in this paper
we shall assume that the geometric background is a flat
Friedmann-Robertson-Walker (FRW), with metric,
\begin{equation}
\label{metricfrw} ds^2 = - dt^2 + a(t)^2 \sum_{i=1,2,3}
\left(dx^i\right)^2\, ,
\end{equation}
with $a(t)$ being the scale factor.

\section{Chern-Simons Modified Axion Dark Matter $f(R)$ Gravity}

The model we propose composes from a vacuum $f(R)$ gravity in the
presence of an axion scalar field, with a CS term present too. The
gravitational action is,
\begin{equation}
\label{mainaction} \mathcal{S}=\int d^4x\sqrt{-g}\left[
\frac{1}{2\kappa^2}f(R)-\frac{1}{2}\partial^{\mu}\phi\partial_{\mu}\phi-V(\phi)+\frac{1}{8}\nu
(\phi)R\tilde{R} \right]\, ,
\end{equation}
where $R\tilde{R}=\epsilon^{abcd}R_{ab}^{ef}R_{cdef}$,
$\kappa^2=\frac{1}{8\pi G}$, with $G$ Newton's gravitational
constant, and $\epsilon^{abcd}$ is the totally antisymmetric
Levi-Civita tensor. Basically the Chern-Pontryagin density is an
analogue of the term $ ^*F_{\mu \nu}F^{\mu \nu}$ built from the
curvature $F_{\mu \nu}$ on a principal bundle with connection
components $A_{\mu}$, but we will call it Chern-Simons term in
order to comply with the literature.

By varying the action (\ref{mainaction}) with respect to the
metric, and by using the FRW metric of Eq. (\ref{metricfrw}), we
obtain the following equations of motion,
\begin{align}\label{eqnsofmkotion}
& 3 H^2F=\kappa^2\frac{1}{2}\dot{\phi}^2+\frac{RF-f+2
V\kappa^2}{2}-3H\dot{F}\, ,\\ \notag &
-3FH^2+2\dot{H}F=\kappa^2\frac{1}{2}\dot{\phi}^2-\frac{RF-f+2
V}{2}+\ddot{F}+2H\dot{F}\, ,
\end{align}
while the variation with respect to the scalar field yields,
\begin{equation}\label{scalarfieldeqn}
\ddot{\phi}+3H\dot{\phi}+V'(\phi)=0\, ,
\end{equation}
where $V'(\phi)=\frac{\partial V}{\partial \phi}$ and
$F=\frac{\partial f}{\partial R}$. As it can be seen from the Eqs.
(\ref{eqnsofmkotion}) and (\ref{scalarfieldeqn}), the CS term does
not affect the background equations, as was also demonstrated by
Hwang and Noh \cite{Hwang:2005hb}, see also \cite{Choi:1999zy}. In
addition, it was shown in \cite{Hwang:2005hb,Choi:1999zy} that the
scalar perturbations remain intact by the presence of the CS-term.
The reason for this is that it is not possible to form a scalar
$T_{00}$ energy momentum tensor nor vector $T_{0 \alpha}$ or
symmetric tensor $T_{\alpha \beta}$, which contains
$\epsilon^{abcd}$ and scalar derivatives only \cite{Choi:1999zy}.
Therefore, the only effect of the CS term is on tensor
perturbations and let us present here the slow-roll indices of the
$f(R)$-scalar theory. We adopt the notation of
\cite{Hwang:2005hb}, and the slow-roll indices are,
\begin{equation}\label{slowrollindices}
\epsilon_1=\frac{\dot{H}}{H^2},\,\,\,\epsilon_2=\frac{\ddot{\phi}}{H\dot{\phi}},\,\,\,\epsilon_3=\frac{\dot{F}}{2HF},\,\,\,\epsilon_4=\frac{\dot{E}}{2HE},\,\,\,\epsilon_6=\frac{\dot{Q}_t}{2HQ_t}\,
,
\end{equation}
where $E$ and $Q_t$ are defined as follows,
\begin{equation}\label{parametersFQe}
E=\frac{F}{\dot{\phi}^2\kappa^2}\left(\dot{\phi}^2+\frac{3\dot{F}^2}{2F\kappa^2}
\right),\,\,\, Q_t=\frac{F}{\kappa^2}+2\lambda_l\dot{\nu}k/a\, .
\end{equation}
The parameter $\lambda_l$ in Eq. (\ref{parametersFQe})
characterizes the polarization of the primordial gravity waves
with wavenumber $k$ and takes values $\lambda_L=-1$ and
$\lambda_R=1$ for left ad right handed polarization states, while
$a$ is simply the scale factor. As it was shown in Ref.
\cite{Hwang:2005hb}, the spectral index of the primordial scalar
curvature perturbations is,
\begin{equation}\label{spectralindexgeneral}
n_s=2\frac{2\epsilon_1-\epsilon_2+\epsilon_3-\epsilon_4}{1+\epsilon_1}\,
,
\end{equation}
while the spectral index of the tensor perturbations is,
\begin{equation}\label{tensorspectralindex}
n_T=2\frac{\epsilon_1-\epsilon_6}{1+\epsilon_1}\, .
\end{equation}
Accordingly, the tensor-to-scalar ratio for the action
(\ref{mainaction}) is equal to,
\begin{equation}\label{tensortoscalarratio}
r=16|\epsilon_1-\epsilon_3|\frac{1}{2}\sum_{l=L,R}\frac{1}{|1+2\lambda_l\frac{\kappa^2k\dot{\nu}}{aF}|}\,
.
\end{equation}
Basically, the effect of the CS term of the axion field can be
seen only on the slow-roll parameter $\epsilon_6$, which affects
the spectral index of the tensor perturbations, and on the
tensor-to-scalar ratio, as was expected, since the background
equations and the slow-roll indices being involved in the scalar
perturbations are not affected. Thus only the tensor perturbations
related phenomenology of $f(R)$ gravity theories is affected by
the presence of the CS term. The aim of this paper is quantify the
impact of the CS term on some characteristic models of $f(R)$
gravity. Prior getting into that, let us briefly discuss the
essential phenomenology of axion dark matter theories. A recent
review on axion cosmology is given in Ref. \cite{Marsh:2015xka}
and we shall adopt the notation and conventions of this review.
The axion field is described by a canonical scalar field with
potential $V(\phi)$, and with the additional assumption that a
string theory inspired CS term is present in the action. The most
promising model of axions that can produce some of the dark matter
present today in our Universe, is the so-called misalignment
model. In the context of misalignment model, the original unbroken
Peccei-Quinn $U(1)$ symmetry is broken and the axion field has a
non-zero vacuum expectation value, which is strongly model
dependent, but quite large, and actually of the order
$\mathcal{O}(10^{19})$GeV or larger. The axion field is at the
broken phase during inflation and the mass of the axion, which we
denote $m_a$ is almost constant, and remains constant until the
radiation domination era. Also during the inflationary era, when
$H\gg m_a$, the axion has a potential,
\begin{equation}\label{axionpotential}
V(\phi(t_i))\simeq \frac{1}{2}m_a^2\phi^2_i(t_i)\, ,
\end{equation}
and the value of the scalar field is actually the constant
non-zero expectation value of the axion. During the inflationary
era, the field is overdamped, and it is practically frozen at its
vacuum expectation value. For all cosmic times during this era, we
have practically the following conditions holding true for the
dynamics of the axion,
\begin{equation}\label{axioninitialconditions}
\dot{\phi}(t_i)=\delta\ll 1,\,\,\,\phi(t_i)=f_a\theta_a\, ,
\end{equation}
where $f_a$ is the axion decay constant, and $\theta_a$ is the
initial misalignment angle. Thus during inflation, the axion
effective equation of state is approximately $w_{eff}=-1$ and it
basically acts as a cosmological constant term in the cosmological
equations, since the potential is constant. Hence, the energy
density of the axion at early times and before it starts
oscillations, is solely determined by the axion mass and by the
initial field displacement $\phi_i$. Practically the axion field
is very slowly varying, so that $\dot{\phi}=\delta\ll 1$ and also
$\ddot{\phi}=\zeta \ll 1$ during the whole inflationary era, until
it starts to oscillate during the radiation domination era. We
shall assume that low-scale inflation occurs, so by taking into
account the compatibility with the Planck and BICEP2/Keck-Array
data, the inflationary scale is of the order
$H_I=\mathcal{O}(10^{10})$GeV in the low-scale inflation scenario.
Also for phenomenological reasons having to do with overproduction
of axion dark matter via isocurvature perturbations backreaction
and also by taking the allowed axion energy density into account,
the most convenient choices for $f_a$ and $\theta_a$ that fit the
data is \cite{Marsh:2015xka},
\begin{equation}\label{axionconstraint}
f_a\sim \mathcal{O}(10^{11})\mathrm{GeV},\,\,\, \theta_a \sim
\mathcal{O}(1)\, .
\end{equation}
Also the mass of the axion field is assumed to be,
\begin{equation}\label{massaxion}
m_a\sim \mathcal{O}(10^{-12})\mathrm{eV}\, ,
\end{equation}
which is the maximum allowed value in the context of the broken
Peccei-Quinn symmetry scenario. From the form of the equations of
motion (\ref{eqnsofmkotion}) it is obvious that if the $f(R)$
gravity is simply the Einstein-Hilbert gravity $f(R)=R$, the
contribution of the action in the cosmological evolution is
significant, and actually it simply contributes a cosmological
constant $\sim V(\phi)$ in the cosmological equations. However, if
an $f(R)$ gravity is chosen that diverges faster for large
curvatures, then the contribution of the axion can be reduced. In
all cases, the CS term $\nu R\tilde{R}$ affects only tensor
perturbations and not the background evolution. In the following
sections we shall quantify our study by choosing specific $f(R)$
gravity models and we examine in detail their inflationary
phenomenology.

\section{Effects of Chern-Simons Modified Axion Dark Matter on Power-law $f(R)$ and Starobinsky Gravity}

Let us first consider an $f(R)$ gravity with problematic
phenomenology, which is,
\begin{equation}\label{polynomialfr}
f(R)=R+\beta R^n\, ,
\end{equation}
with $n$ being chosen in the interval $n=
\left[\frac{1+\sqrt{3}}{2},2\right]$ in order to have accelerating
expansion. For the moment the approach we shall adopt for the
power-law model, will not cover the Starobinsky model case, which
corresponds to $n=2$, because the Starobinsky model can be treated
in a more accurate way in order to find the exact Hubble rate
realized by the model in the slow-roll approximation. The
differences between the power-law model for $n\neq 2$ and the
Starobinsky model are clarified in the Appendix.

The model (\ref{polynomialfr}) leads to non-zero tensor spectral
index and also one does not have simultaneous compatibility of the
spectral index of primordial scalar perturbations and of the
tensor-to-scalar ratio with the observational data. In this
section we shall investigate the phenomenological implications of
the CS term and of the presence of the dark matter axion field in
the theory. Firstly, by recalling that during the inflationary era
$\dot{\phi}$ is constant and $\dot{\phi}=\delta \ll 1$, and also
that the scalar field is almost constant, see the conditions
(\ref{axioninitialconditions}), the first equation of motion in
Eq. (\ref{eqnsofmkotion}) becomes,
\begin{align}\label{eqnsofmkotionfrpoly}
& 3 H^2n\beta R^{n-1}=\frac{\kappa^2\delta^2}{2}+\frac{\beta
(n-1)R^{n-1}}{2}-3n(n-1)\beta
HR^{n-2}\dot{R}+\frac{\kappa^2}{2}m_a^2f_a^2\theta_a^2\, ,
\end{align}
so by using the fact that at $R=12H^2+6\dot{H}$, we get at leading
order,
\begin{equation}\label{leadingordereqn}
3H^2n\beta \simeq 6\beta (n-1)H^2-6n\beta(n-1)\dot{H}+3\beta
(n-1)\dot{H}+\kappa^2\frac{1}{2
(12H^2)^{n-1}}\delta^2+\frac{\kappa^2}{2
(12H^2)^{n-1}}m_a^2f_a^2\theta_a^2\, .
\end{equation}
The last two terms in Eq. (\ref{leadingordereqn}) are much smaller
than the rest of the terms, therefore these can be neglected. In
order to have a concrete idea on how small are the last two terms
in Eq. (\ref{leadingordereqn}), let us use the numerical values
for the axion field and for the low-scale inflation we presented
in the previous section. The first term is obviously suppressed
due to the presence of $\kappa^2\delta^2$ in the numerator and
$(12H^2)^{n-1}$ in the denominator, so let us focus on the last
term in Eq. (\ref{leadingordereqn}). Recall that $n=
\left[\frac{1+\sqrt{3}}{2},2\right)$, so let us use the values for
$m_a$, $f_a$ and $\theta_a$ appearing in Eqs.
(\ref{axionconstraint}) and (\ref{massaxion}), and also we assume
that the Hubble rate during inflation is
$H_I=\mathcal{O}(10^{10})$GeV. By using these values and also
since $\kappa^2=4.10\times 10^{-28}$eV, we get that the last term
is of the order,
\begin{equation}\label{termlastleading}
\frac{\kappa^2}{2
(12H^2)^{n-1}}m_a^2f_a^2\theta_a^2=\mathcal{O}(10^{-70}/\beta)eV\,
,
\end{equation}
while the first term in Eq. (\ref{leadingordereqn}) is of the
order $\mathcal{O}(10^{38})$eV. Hence, due to the fact that in
most polynomial inflationary $f(R)$ gravities, $\beta\ll 1$, it is
obvious that the last two terms are subleading for the axion dark
matter $f(R)$ gravity model, so these can be safely neglected.
Therefore, the differential equation that determines the Hubble
rate is,
\begin{equation}\label{leadingordereqn}
3H^2n\beta \simeq 6\beta (n-1)H^2-6n\beta(n-1)\dot{H}+3\beta
(n-1)\dot{H}\, ,
\end{equation}
which can be solved to yield,
\begin{equation}\label{hubblefrpoly}
H(t)=\frac{-2n^2+3n-1}{(n-2)t}\, .
\end{equation}
Using the Hubble rate (\ref{hubblefrpoly}), the slow-roll indices
$\epsilon_i$, $i=1,..,4$ can be easily calculated for the $f(R)$
gravity (\ref{polynomialfr}), at the horizon crossing time
instance $t_k$ during inflation, and these are,
\begin{equation}\label{epsiloniforfrpoly}
\epsilon_1=\frac{n-2}{1-3n+2n^2},\,\,\,\epsilon_2\simeq
0,\,\,\,\epsilon_3=(n-1)\epsilon_1,\,\,\,\epsilon_4=\frac{n-2}{n-1}\,
,
\end{equation}
where we used the fact that $\dot{\phi}(t_k)=\delta\ll 1$ and
$\ddot{\phi}(t_k)=\zeta\ll \delta \ll 1$, so $\epsilon_2\simeq 0$.
The slow-roll indices (\ref{epsiloniforfrpoly}) have the same
functional form as the vacuum $f(R)$ gravity, however the
slow-roll index $\epsilon_6$ contains the effects of the CS axion
term, and now we shall find its explicit functional form. From
Eqs. (\ref{slowrollindices}) and (\ref{parametersFQe}), we have,
\begin{equation}\label{epsilon6explict}
\epsilon_6=\frac{1}{2}\sum_{l=L,R}\left(
\frac{\dot{F}}{2HF}+\frac{\lambda_l\dot{x}\kappa^2}{2FH}\right)\left(\frac{1}{1+\frac{\lambda_lx\kappa^2}{F}}
\right)\, ,
\end{equation}
where $x=\frac{2\dot{\nu}k}{a}$, $F\simeq n\beta R^{n-1}$. The
slow-roll parameter $\epsilon_6$ can be rewritten,
\begin{equation}\label{epsilon6explict}
\epsilon_6=\frac{1}{2}\sum_{l=L,R}\left(
\epsilon_3+\frac{\lambda_l\dot{x}\kappa^2}{2FH}\right)\left(\frac{1}{1+\frac{\lambda_l
x\kappa^2}{F}} \right)\, .
\end{equation}
After performing the sum over the polarization states, and by
taking into account that $\lambda_L=-1$ and $\lambda_R=1$, Eq.
(\ref{epsilon6explict}) can be rewritten as follows,
\begin{equation}\label{polarizedepsilon6}
\epsilon_6=\frac{\epsilon_3}{2}\left(\frac{1}{1-\frac{\kappa^2x}{F}}+\frac{1}{1+\frac{\kappa^2x}{F}}
\right)+\frac{1}{2}\sum_{l=L,R}\frac{\lambda_l\kappa^2\dot{x}}{2FH}\left(\frac{1}{1+\frac{\lambda_l\kappa^2x}{F}}\right)\,
.
\end{equation}
In addition, the tensor-to-scalar ratio for the model at hand is
equal to,
\begin{equation}\label{cscorrtensortoscalar}
r=16
|\epsilon_1-\epsilon_3|\frac{1}{2}\left(\frac{1}{|1-\frac{\kappa^2x}{F}|}+\frac{1}{|1+\frac{\kappa^2x}{F}|}
\right)\, .
\end{equation}
At this point we can obtain some phenomenological results for the
CS axion corrected power-law $f(R)$ gravity model. In order to
have a spectral index of scalar primordial perturbations with
value $n_s=0.965$, we easily find from Eq.
(\ref{spectralindexgeneral}), we must have $n=1.817$. So the
simple power-law $f(R)$ gravity model without the CS axion
corrections yields a tensor-to-scalar ratio $r_v=0.24$, which is
excluded from both Planck \cite{Ade:2015lrj} and BICEP2/Keck-Array
data \cite{Array:2015xqh}. The CS corrected tensor-to-scalar ratio
can be compatible with the observations, due to the presence of
the $x/F$ terms. Particularly, suppose that we need to achieve a
value compatible with the BICEP2/Keck-Array data, so assume that
$r=0.06$. This can be achieved if $\frac{\kappa^2x}{F}=4.37$, and
in addition, if $\frac{\kappa^2x}{F}$ is chosen to take larger
values, the tensor-to-scalar ratio takes even smaller values, for
example if $\frac{\kappa^2x}{F}=24.93$, then $r=0.01$. Thus the
phenomenology of the power-law $f(R)$ gravity model
(\ref{polynomialfr}) becomes refined by the presence of the CS
corrected axion dark matter.

Now let us turn our focus on the tensor spectral index $n_T$,
which in the vacuum $f(R)$ gravity case can be found from Eqs.
(\ref{tensorspectralindex}) and (\ref{polarizedepsilon6}) (for
$x=0$) that it is equal to,
\begin{equation}\label{ntclassical}
|n_T|=0.03\, ,
\end{equation}
which is non-zero and thus the theory is problematic. However as
we shall show, the tensor spectral index can be significantly
smaller in the context of CS corrected axion $f(R)$ gravity case.
This can be seen from Eq. (\ref{epsilon6explict}), due to the
presence of the $x$-dependent terms. The result is strongly model
dependent, so let us choose the function $\nu (\phi)$ to be,
\begin{equation}\label{nuphipowerlaw}
\nu (\phi)=\Lambda e^{\kappa \phi}
\end{equation}
where $\Lambda$ a free parameter, and $\dot{\nu}$, $\ddot{\nu}$
are equal to,
\begin{equation}\label{dotnu}
\dot{\nu}\simeq \Lambda\kappa e^{\kappa
\phi}\delta,\,\,\,\ddot{\nu}\simeq
\Lambda\kappa^2\delta^2e^{\kappa \phi}\, .
\end{equation}
There is no specific reason for choosing the $\nu (\phi)$ function
in Eq. (\ref{nuphipowerlaw}), apart from the fact that the
derivatives of $\nu (\phi)$ with respect to the cosmic time
contain the same exponential term $\sim e^{\kappa \phi}$ so a
direct comparison between the resulting terms may lead to a
conclusion on which is dominant at leading order. One can choose
the function $\nu (\phi)$ freely for the moment, because there is
no direct proof of the existence of the axion, or the presence of
the string axion coupling. However, the observation of axions and
also the presence of non-equivalent polarizations in the
primordial gravitational waves may provide sufficient data to find
the functional form of the function $\nu (\phi)$. Thus for the
moment we use the exponential form only for simplicity.

Having Eq. (\ref{dotnu}) at hand we can evaluate $\epsilon_6$ at
the horizon crossing time instance $t=t_k$, when $k=Ha$, and
$\phi_k=\theta_af_a$, so we have,
\begin{equation}\label{firstterm1}
\frac{\kappa^2x}{F}\simeq \frac{\delta  \kappa ^3 \Lambda 2^{3-2
n} 3^{1-n} H_I^{2-n} e^{f_a \theta_a \kappa }}{\beta  n}\, ,
\end{equation}
and in addition,
\begin{equation}\label{secondterm}
\frac{\kappa^2\dot{x}}{2FH}=\frac{\delta ^2 \kappa ^5 \Lambda
12^{1-n} H_I^{1-n} e^{f_a \theta_a \kappa }}{\beta n}-\frac{c_a
\delta  \kappa ^3 \Lambda  12^{1-n} H_I^{1-n} e^{f_a \theta_a
\kappa }}{\beta n}\, ,
\end{equation}
where $c_a$ is equal to,
\begin{equation}\label{ca}
c_a=\frac{n-2}{-2 n^2+3 n-1}\, .
\end{equation}
Basically, the term $\frac{\kappa^2\dot{x}}{2FH}$ is subleading in
Eq. (\ref{polarizedepsilon6}), as we now demonstrate. Indeed, by
taking the values for the free parameters as in the previous
section, we have,
\begin{equation}\label{xf1}
\frac{\kappa^2x}{F}=\frac{4.165985637444943 \times 10^{-80} \delta
\Lambda }{\beta }\, .
\end{equation}
So in order to have $r=0.01$, we must have
$\frac{\kappa^2x}{F}=24.93$, thus from Eq. (\ref{xf1}) the
parameters $\beta$, $\Lambda$ and $\delta$ must satisfy,
\begin{equation}\label{parms}
\frac{\beta}{\delta \Lambda}\simeq 1.73 \times 10^{-81}\, ,
\end{equation}
thus we have approximately,
\begin{equation}\label{finalsecondterm}
\frac{\frac{\kappa^2\dot{x}}{2FH}}{\frac{\kappa^2x}{F}}\simeq
-1.07436\times 10^{-19}+4.9281\times 10^{-46}\delta\, ,
\end{equation}
and since $\delta\ll 1$, the second term in Eq.
(\ref{polarizedepsilon6}) is significantly suppressed, so we have
approximately,
\begin{equation}\label{polarizedepsilon6approximatefinal}
\epsilon_6\simeq
\frac{\epsilon_3}{2}\left(\frac{1}{1-\frac{\kappa^2x}{F}}+\frac{1}{1+\frac{\kappa^2x}{F}}
\right)\, .
\end{equation}
This is the reason we chose the exponential function $\nu (\phi)$,
in order to have a clear picture of which terms are dominant, due
to the presence of the same term $\sim e^{\kappa \phi}$ in the
derivatives of $\nu (\phi)$ with respect to the cosmic time.

Therefore, for $\frac{\kappa^2x}{F}\simeq 24.93$, we have
$|n_T|\simeq 0.000116871$, which is $10^{-3}$ times smaller in
comparison to the vacuum $f(R)$ gravity result of Eq.
(\ref{ntclassical}). Thus it is clear that the CS axion dark
matter $f(R)$ gravity results to better phenomenological results,
in comparison to the vacuum $f(R)$ gravity case. Of course, the
results are model dependent and considerable fine-tuning is
required in order to obtain refined phenomenological results, but
the general outcome is phenomenologically more appealing in
comparison to the axion free $f(R)$ gravity, at least for the
polynomial $f(R)$ gravity of Eq. (\ref{polynomialfr}). Also we
need to mention that the presence of $F$ in the denominator of the
term $\frac{\kappa^2\dot{x}}{2FH}$ requires less fine tuning in
order to achieve a desirable value for
$\frac{\kappa^2\dot{x}}{2FH}$ in comparison to the
Einstein-Hilbert considerations of Ref. \cite{Choi:1999zy}, in
which case $F=1$. The fine-tuning is unavoidable for the moment in
order to see how the theory can be fit to the present
observational data. Indeed, we have two sources of uncertainty in
the present outcome, firstly the $f(R)$ gravity model, and
secondly the axion coupling $\nu (\phi)$. The purpose of this
example is to show the new possibilities that arise in $f(R)$
gravity phenomenology by the presence of a CS axion coupling, and
the result is that even non-viable vacuum $f(R)$ gravity models
may become compatible with the observations. In the future, the
observational data will possibly indicate if some model of $f(R)$
gravity is indeed the correct description for inflation, and of
course if the axion exists. In this case, one may severely
constrain the free parameters of $f(R)$ gravity, in the case at
hand $\beta$, and also find hints for the presence (or non
presence) of the axion CS coupling. Hence, these fine-tunings will
be less severe, and of course more physically motivated.

In the context of CS axion dark matter corrected $f(R)$ gravity,
it is also possible to make a viable $f(R)$ gravity to have even
smaller primordial gravitational radiation. In view of this
aspect, we now discuss the case of Starobinsky inflation
\cite{Starobinsky:1982ee}. In this case, the $f(R)$ gravity is of
the form,
\begin{equation}\label{starobinsky}
f(R)=R+\frac{1}{36H_i}R^2\, ,
\end{equation}
and the Friedman equation in the presence of the misalignment
axion is,
\begin{equation}\label{patsun}
\ddot{H}-\frac{\dot{H}^2}{2H}+3H_iH=-3H\dot{H}\, .
\end{equation}
The above result is due to the fact that the kinetic term of the
axion scalar field and of the corresponding scalar potential are
significantly suppressed during the inflationary era, and the
axion is ``frozen'' in its vacuum expectation value. In order to
have a concrete idea on how much suppressed are these terms, the
potential term over $F$ is in this case,
\begin{equation}\label{termlastleading}
\frac{\kappa^2}{2
(12H^2)}m_a^2f_a^2\theta_a^2=\mathcal{O}(10^{-39}/\beta)eV\, ,
\end{equation}
where we took into account that the inflationary scale $H_I$ in
this case is $H_I=\mathcal{O}(10^{13})$GeV and we used the
previous conventions for the axion field.

The differential equation (\ref{patsun}) can easily be solved, and
it yields the quasi-de Sitter evolution,
\begin{equation}\label{quasidesitter}
H(t)=H_0-H_i t\, ,
\end{equation}
and due to the fact that the slow-roll indices in the Starobinsky
inflation case satisfy $\epsilon_i\ll 1$, the spectral index of
the primordial scalar perturbations and the tensor-to-scalar ratio
in the presence of the axion field are at leading order,
\begin{equation}\label{spectralstarobinsky}
n_s=1-\frac{2}{N},\,\,\,r\simeq
\frac{r_s^v}{2}\left(\frac{1}{|1-\frac{\kappa^2x}{F}|}+\frac{1}{|1+\frac{\kappa^2x}{F}|}\right)\,
,
\end{equation}
where $r_s^v= 48\epsilon_1^2$ is the vacuum $f(R)$ gravity
tensor-to-scalar ratio. The presence of the term $\sim
\frac{\kappa^2x}{F}$ can further reduce the value of the
tensor-to-scalar ratio below the vacuum $f(R)$ gravity value which
is $r_s^v=0.0033$ which is obtained for $N=60$ $e$-foldings. For
example if $\frac{\kappa^2x}{F}=\mathcal{O}(3\times 10^2)$, the
tensor-to-scalar ratio is $r=\mathcal{O}(10^{-5})$, while for
large values, for example $\frac{\kappa^2x}{F}=\mathcal{O}(3\times
10^{8})$, the tensor-to-scalar ratio is $r=\mathcal{O}(10^{-11})$.
Finally let us discuss the effect of the CS axion coupling on the
tensor spectral index $n_T$. In the vacuum Starobinsky model, this
is exactly equal to zero, however in the presence of the CS axion
coupling term, the tensor spectral index is not equal to zero
anymore, due to a non-trivial $\epsilon_6$ slow-roll index. In
this case, a similar analysis as in the previous case indicates
that $n_T\sim 0$, when $\frac{\kappa^2x}{F}\gg 1$, so this means
that the Starobinsky inflation model in the presence of the CS
axion dark matter has the same spectral index as the vacuum $f(R)$
gravity model, but many orders reduced tensor-to-scalar ratio, and
thus it produces smaller amounts of inflationary gravitational
radiation.

Before closing this section, let us briefly discuss an interesting
perspective of CS axion dark matter $f(R)$ gravity models. In the
context of the misalignment axion dark matter, during the
inflationary era the axion field is frozen at its vacuum
expectation value, and thus affects inflation via the tensor
perturbations quantified by the CS coupling function $\eta
(\phi)$. Thus $f(R)$ gravity drives inflation, at least the
background evolution and also controls the primordial scalar
perturbations, however the axion affects the tensor perturbations,
reducing the amount of primordial gravity waves. As the Hubble
rate drops, and specifically when $H\sim m_a$, the axion field
starts to oscillate, and thus this could be viewed as a natural
graceful exit mechanism and also some reheating type. When the
Hubble rate satisfies $H\ll m_a$, the curvature is small and thus
the axion field starts to dominate the cosmological evolution. The
WKB approximation in the context of standard Einstein-Hilbert
gravity yields a solution $\phi (t)\sim \cos (m_a t+ \theta)$, so
if indeed the $f(R)$ gravity does not control the cosmological
evolution during this era, this WKB extracted solution will be
indeed a solution to the scalar equation of motion. This yields an
energy density for the axion  $\rho\sim a^{-3}$ and also an
averaged axion effective equation of state parameter $\langle
w_{eff}\rangle\sim 0$, for $t\gg 1/m_a$, independently of the
background being radiation or matter dominated
\cite{Marsh:2015xka}. Thus if one finds a model of $f(R)$ gravity
which dominates at early times, like the Starobinsky model, and
remains subdominant at intermediate evolutionary stages of the
Universe, while it dominates at late times again, then the CS
axion dark matter $f(R)$ gravity could potentially provide an
appealing unified description of early and late-time acceleration
eras, with the intermediate eras. Here we just sketched a
qualitative picture however, but this issue is quite interesting
for future development.

\section{Discussion and Concluding Remarks}

The possibility that the axion could be the main constituent of
dark matter is stimulating, and the detection of the axion is one
of the main goals in several observational and experimental
proposals \cite{Du:2018uak,Henning:2018ogd,Ouellet:2018beu}.
Actually the misalignment models which are based on small mass
axions, could provide a major candidate for dark matter, and in
the very appealing experimental proposals of Refs.
\cite{Du:2018uak,Henning:2018ogd,Ouellet:2018beu}, the search for
low mass axions is the main aim. Also axions can have indirect
effects to neutron stars via their interaction with the thermal
photons near the core of neutron stars, and several observational
proposals exist \cite{Safdi:2018oeu}, see also
\cite{Caputo:2018vmy,Caputo:2018ljp}, that could actually verify
the existence of axions in the near future. Actually axions could
interact with photons \cite{Balakin:2009rg} and electric fields in
a plasma \cite{Balakin:2012up,Balakin:2014oya} and therefore
neutron stars could be a virtual future laboratory for seeking
axion induced effects. In the literature there exist several
theoretical proposals discussing the possibility of detecting
axions, see for example \cite{Avignone:2018zpw}. Thus if the axion
is one or the main components of dark matter, this could be
revealed in the near future.

One of the issues we would like now to briefly discuss is the
implications of the CS axion term $\nu (\phi)\tilde{R}R$ on the
propagation of gravitational waves. Particularly, it is known for
quite some time \cite{Hwang:2005hb} that the presence of the CS
term discriminates the two different polarizations of the
primordial gravity waves. In the literature this is a well studied
possibility \cite{Inomata:2018rin,Kamionkowski:1997av}, an effect
which is known as parity violating gravity waves, and this
non-equivalence in the propagation of gravity waves could have an
observable effect on the Cosmic Microwave Background
\cite{Pritchard:2004qp}, which could be captured in the next
generation of experiments. In this paper we also demonstrated that
the presence of the CS term can reduce significantly the
tensor-to-scalar ratio of $f(R)$ gravity theories. We specified
our analysis by using two concrete examples, a power-law $f(R)=
R+\beta R^n$ gravity (apart of realistic $n=2$ case which
corresponds to Starobinsky inflation) and also the $R^2$ gravity.
In the case of power-law gravity ($n<2$), the vacuum theory was
unable to provide a phenomenologically viable theory, however the
presence of CS corrected axion dark matter can modify the
resulting theory, suppressing the tensor perturbations and the
corresponding tensor-to-scalar ratio. In the context of $f(R)$
gravity, the tensor-to-scalar is suppressed due to the presence of
the term $\frac{\kappa^2k\dot{\nu}}{aF}$. In the case of
Einstein-Hilbert gravity, the term $F=\frac{\partial f}{\partial
R}$ is equal to one, however in the case of $f(R)$ gravity, it is
proportional to powers or functions of the curvature, hence the
suppression of the term $\frac{\kappa^2k\dot{\nu}}{a}$ caused by
the $\kappa^2$ in the Einstein-Hilbert case, is not an issue
anymore in the context of $f(R)$ gravity. Finally, with regard to
the Starobinsky case in the context of CS axion dark matter
gravity, we demonstrated in a quantitative way that the resulting
inflationary theory could have significantly smaller
tensor-to-scalar ratio in comparison to the vacuum theory. Thus,
the CS axion dark matter could extend the viability of already
viable $f(R)$ gravity cosmological models, in a way that the
scalar perturbations and the background evolution are unaffected,
but only the gravitational radiation is affected. Finally, we
should briefly note that the axion sinusoidal oscillations when
$H\sim m_a$ could actually trigger the graceful exit from
inflation even in the context of polynomial $f(R)$ gravity, which
can be problematic in the vacuum theory, and also can contribute
to the $f(R)$ gravity reheating mechanism. We hope to address this
issue in more detail in a future work more focused on this issue,
but this study should require a numerical approach.

\section*{Appendix: Power-law and Starobinsky $f(R)$ Gravity Models in the slow-roll Approximation}

In this section we shall demonstrate the differences in deriving
the cosmological evolution stemming from the polynomial $f(R)$
gravity model of Eq. (\ref{polynomialfr}) for $n\neq 2$ and for
the Starobinsky model (\ref{starobinsky}). As we now show, the
Starobinsky model leads to less complicated and more accurate
results, due to the fact that $n=2$. We shall consider the vacuum
case, since the scalar field contribution will be neglected
eventually. The first Friedman equation for the vacuum $f(R)$
gravity is,
\begin{equation}\label{friedmannewappendix}
3 H^2F=\frac{RF-f}{2}-3H\dot{F}\, .
\end{equation}
Let us first consider the polynomial $f(R)$ gravity of Eq.
(\ref{polynomialfr}), in which case, by assuming that $F\sim
n\beta R^{n-1}$ the Friedman equation (\ref{friedmannewappendix})
becomes,
\begin{align}\label{eqnsofmkotionfrpolyappendix}
& 3 H^2n\beta R^{n-1}=\frac{\beta (n-1)R^{n-1}}{2}-3n(n-1)\beta
HR^{n-2}\dot{R}\, ,
\end{align}
so by using the fact that at $R=12H^2+6\dot{H}$,  and by further
taking the simplification $R\sim 12 H^2$ and $\dot{R}\sim
24H\dot{H}$, the Friedman equation
(\ref{eqnsofmkotionfrpolyappendix}) becomes approximately at
leading order,
\begin{equation}\label{leadingordereqnappendix}
3H^2n\beta \simeq 6\beta (n-1)H^2-6n\beta(n-1)\dot{H}+3\beta
(n-1)\dot{H}\frac{1}{2 (12H^2)^{n-1}}\delta^2\, .
\end{equation}
So by simplifying the above we get,
\begin{equation}\label{leadingordereqnappendix}
3H^2n\beta \simeq 6\beta (n-1)H^2-6n\beta(n-1)\dot{H}+3\beta
(n-1)\dot{H}\, ,
\end{equation}
which can be solved to yield,
\begin{equation}\label{hubblefrpolyappendix}
H(t)=\frac{-2n^2+3n-1}{(n-2)t}\, .
\end{equation}

Let us now consider the Starobinsky model case,
\begin{equation}\label{starobinskyappendix}
f(R)=R+\frac{1}{36H_i}R^2\, ,
\end{equation}
so by substituting $R=12H^2+6\dot{H}$ and
$\dot{R}=24H\dot{H}+6\ddot{H}$ and  $F=1+\frac{R}{18H_i}$ in the
Friedman equation (\ref{friedmannewappendix}), we obtain exactly
the following differential equation,
\begin{equation}\label{patsunappendix}
\ddot{H}-\frac{\dot{H}^2}{2H}+3H_iH=-3H\dot{H}\, .
\end{equation}
The first two terms can be disregarded during the slow-roll era,
so the resulting differential equation is,
\begin{equation}\label{patsunappendix1}
3H_iH=-3H\dot{H}\, ,
\end{equation}
which when solved yields the quasi-de Sitter evolution of Eq.
(\ref{quasidesitter}). So basically in the Starobinsky model,
there is no fractional power of $n$ complicating things, and we
take $R=12H^2+6\dot{H}$ and $\dot{R}=24H\dot{H}+6\ddot{H}$. The
only simplification assumed is the slow-roll condition in Eq.
(\ref{patsunappendix}). In the power-law case we had to do three
stages of simplifications in order to extract an analytic result,
so it is a more complicated case.

\section*{Acknowledgments}

This work is supported by MINECO (Spain), FIS2016-76363-P, and by
project 2017 SGR247 (AGAUR, Catalonia) (S.D.O).

\end{document}